\documentclass[twocolumn,twoside,slac]{revtex4}
\usepackage{graphicx}
\usepackage{fancyhdr}
\begin{document}

\title{Measures of Significance in HEP and Astrophysics}


\author{James T. Linnemann}
\affiliation{Michigan State University, E. Lansing, MI 48840, USA}
\affiliation{Los Alamos National Laboratory, Los Alamos, NM 87545,
USA}

\begin{abstract}
I compare and discuss critically several measures of statistical
significance in common use in astrophysics and in high energy
physics. I also exhibit some relationships among them.

\end{abstract}

\maketitle

\thispagestyle{fancy}


\section{INTRODUCTION}
Significance testing for a possible signal in counting experiments
centers on the probability that an observed count in a signal
region, or one more extreme, could have been produced solely by
fluctuations of the background source(s) in that region.
Statisticians refer to this probability as a p-value. The
traditions for calculating signal significance differ between
High Energy Physics (HEP) and High Energy Gamma Ray Astrophysics
(GRA).  Both fields often quote significances in terms of
equivalent standard deviations of the normal distributions
(statisticians sometimes refer to this as a Z-value).

I will present several of the commonly used methods in HEP and
GRA, apply them to examples from the literature, then discuss the
results. Here I will concentrate on observed significance, the
significance of a particular observation, rather than predictions
of significance for a given technique as a function of exposure.
The prediction problem is slightly different, involving the power
of the test, or the probability of making an observation at a
given significance level.

GRA has emphasized simple, quickly-evaluated analytical formulae
for calculating Z directly (choosing asymptotically normal
variables), while HEP has typically calculated probabilities
(p-values) and then translated into a Z-value by

$p = P(s \geq \mathrm{observed}\ |\  \mathrm{assume\ only\
background}\ );$

$ Z = \Phi^{-1}(p);\ \   \Phi(z) = \frac{1}{\sqrt{2 \pi}} \int
_{-\infty }^{z} \
e^{{-t^2}\big/2} \ d t $\\
This relation can be written\cite{abro} for large $Z
> 1.5$ as

$Z \approx \sqrt{u - Ln\ u} ;\ \ u = - 2 Ln (p \sqrt{2 \pi} ) $\\
giving a rough dependence of $Z \sim \sqrt{- Ln\ p}$.  While more
general than the search for a simple formula for Z-values, the HEP
approach loses track of the analytic structure of the problem.

Observations in GRA typically consist of a count of gamma rays
when pointing directly at a potential source, called an on-source
count, $N_{on}$.  The analogous quantity in HEP is the number of
counts in a signal region. The background relevant to an
observation of a source is typically estimated in GRA by an
off-source observation.  The relative exposure of the two
observations is denoted by $\alpha = T_{on}/T_{off}$, often less
than unity. Then the background count mean's estimate is $b\ =
\alpha N_{off}$, its (Poisson) uncertainty $\delta b = \alpha
\sqrt{N_{off}}$, and thus one derives
\begin{equation}\label{eq:alpha}
\alpha =(\delta b)^2 / b
\end{equation}
GRA expressions are couched in terms of $\alpha$.
%In what follows,
I will also use $x = N_{on},\ y = N_{off},\ k=x+y $ for
compactness.

In HEP, sometimes a side-band method of background estimation is
used, rather like in a GRA measurement; or $b$ may be estimated
as a sum of contributions from Monte Carlo and data-based
side-band estimates, so that often $b \pm \delta b$ is quoted,
where $\delta b$ is derived from adding uncertainties in
quadrature. One can use Eq.\ref{eq:alpha} to \textit{define}
$\alpha$ when comparing HEP results with GRA expressions.
Non-integer values for effective $N_{off}$ result, but usually
cause no problems.

\section{Z-VALUE VARIABLES}
Many expressions for Z are of the form of a ratio of estimates of
signal to its variance, where the signal is estimated by $s =
N_{on} - b = x - \alpha y$.  Then $Z = s/\sqrt{V}$, where  $V$ is
a variance estimate for $s$. A standard GRA reference\cite{li}
gives as an example (their Equation 5) $V_5 = N_{on} + \alpha ^2
N_{off}$. The authors note that this expression treats $N_{on}$
and $N_{off}$ as independent; this does not consistently calculate
$V$ under the null hypothesis, $\mu_{on} = \alpha \mu_{off}$ and
in fact biases against signals for $\alpha < 1$ by overestimating
$V$.  I have derived a related formula, $V_5\prime =
\alpha(1+\alpha)N_{off}$, by using only the background to
estimate the mean and variance: while not optimal, it at least is
consistent with the null.   They also provide $V_9 =\alpha
(N_{on} + N_{off})$, which better implements the null hypothesis.
However, their widely-used recommendation is likelihood ratio
$L(\mu_s,\mu_b)/L(\mu_b)$,

$Z_L = \sqrt{2}\  (\ x\ Ln \frac{x (1+\alpha)}{k \alpha} + y\ Ln
\frac{y (1+\alpha)}{k}\ )^{\frac{1}{2}}$.\\
$Z_L$ derives from the standard likelihood ratio test for a
composite hypothesis, and Wilks' Theorem, giving its asymptotic
normal behavior.  The numerator and denominator likelihoods are
each separately maximized: one for a signal + background model,
the other for a background-only (null) model.

One may instead seek an asymptotically normal variable with
nearly constant variance\cite{zhang},

$Z_{0} = \frac{2}{\sqrt{1 + \alpha} } ( \sqrt{x + 3/8} -
\sqrt{\alpha(y
+ 3/8)}\ )$.\\
The $3/8$ speeds convergence to normality from the underlying
discreteness.

\subsection{Other Frequentist Methods}
One widely used form is $Z_{sb} = s/\sqrt{b}$
(sometimes\cite{astrostats} called the ``signal to noise ratio'').
This entirely ignores the uncertainty in the background estimate.
It is often used for optimizing selection criteria, because of its
simplicity. Slightly better is a $Z_P$ calculated from the
Poisson probability p-value:

$p_P = P(\geq x|b) = \sum_{j=x}^{\infty} e^{-b}b^j/j!=\Gamma (x,0,b)/\Gamma  (x)$. \\
here written\cite{ord} in terms of an incomplete $\Gamma$
function. $Z_P$ still ignores uncertainty in $b$. Occasionally
one sees substitutions of $b \rightarrow b + \delta b$ as a
feeble attempt to incorporate the uncertainty in b.

Finally, one may view a significance calculation directly as a
p-value calculation which one could use as a test of the null
hypothesis. $Z_L$ use the standard (non-optimal) test of a
composite hypothesis against a null.  However, the relationship
of the Poisson means, whether $\mu_{on} > \alpha \mu_{off}$, is a
special case of a composite hypothesis test that admits a more
optimal solution. There exists a Uniformly Most Powerful test
among the class of Unbiased tests for this case, in the form of a
binomial proportion test for the \textit{ratio} of the two Poisson
means\cite{lehman}.  The UMPU properties are, strictly speaking,
derived only with an assumption of randomization, that is, hiding
the underlying discreteness by adding a random number to the
data.  This test yields a binomial probability p-value (using
$k=x+y$):

$p_{Bi} = P_{Bi}(\geq x |\ w, k)  =
 \sum_{j=x}^{k} \frac{k!}{j!(k-j)!}w^j(1-w)^{k-j}$,\\
where $w = \alpha/(1+\alpha)$ is the expected ratio of the
Poisson means for $x$ and $x+y$.  After some manipulation, this
can be written in terms of incomplete and complete beta
functions\cite{ord,abro}, which is convenient for numerical
evaluation:

$p_{Bi} = B(w,x,1+y)/B(x,1+y)$\\
%an alternative representation\cite{zhang} in terms of the $F$
%distribution is also available.
%
This  test is conditional on $x+y$ fixed because of the existence
of a nuisance parameter: there are two Poisson means, but the
quantity of interest is their ratio. While this test is known to
both the GRA\cite{zhang} and HEP\cite{james} communities, it is
common practice in neither, and its optimality properties are not
common knowledge.

Given the (restricted) optimality of the test, and the lack of a
 UMP test for this class of composite hypotheses, this
test ought to be more frequently used to calculate significance,
even though it is clearly a longer calculation than $Z_L$.  For
moderate $x,\ y$, closed forms in terms of special functions are
available, while some care is required for larger $n$.  For $Z_B
< 3$, the Z-values reported may be somewhat too
small\cite{dag,zhang}, but for typical applications one is more
interested in $Z_B > 4$.

It is interesting to note that taking a normal approximation to
the binomial test (that is, comparing the difference of binomial
proportion from its expected value, to the square root of its
normal-approximation variance) yields $(x/k -
w)/\sqrt{w(1-w)/k}$, which can be shown to be identical to $Z_9 =
s/V_9$.

A different approach attempts to move directly from likelihood to
significance by using a 3rd-order expansion\cite{fraser}.  The
mathematics is interesting, combining two first order estimates
(which give significance to order $1/\sqrt{n}$) to yield a
$1/\sqrt{n^3}$ result.  Typically, the first-order estimates are
of the form of a normal deviation, $Z_t$ (like $Z_9$), and a
likelihood ratio like $Z_L$; of these, the likelihood ratio is
usually a better first-order estimate.  The two are then
combined into the third order estimate by a formula such as\\
$Z_3 =Z_L + \frac{1}{Z_L} Ln (Z_t/Z_L)$. \\
Generically, $Z_t = \Delta/\sqrt{V}$ is a Student t-like variable,
where $\Delta$ is the difference of the maximum likelihood value
of $\theta$ (the parameter of interest) from its value under the
null hypothesis, and $V$ is a variance estimate derived from the
Fisher Information $\partial ^2 L/\partial ^2 \theta$.  The
attraction of the method is to achieve simple formulae with
accurate results.  However, the mathematics becomes more
complex\cite{reid} when nuisance parameters are included, as is
needed when the background is imperfectly known. Here I will only
compare the approximate calculation for a perfectly known
background to the corresponding exact calculation, $p_P$.

\section{BAYESIAN METHODS}
HEP common practice often involves Bayesian methods of
incorporating ``systematic'' uncertainties for quantities such as
efficiencies\cite{CH}.  These methods are also used for
calculating significances, particularly when the background $b$
is a sum of several contributions, since the method naturally
extends to complex situations where components of $\delta b$ are
correlated.  The typical calculation represents the lack of
knowledge of $b$ by a posterior density function $p(b|y)$; it is
referred to as a posterior density because it is posterior to the
off-source measurement $y$.  The usual way of proceeding is to
calculate Poisson p-values $p_P = P(\geq x|b)$ as was done above,
but this time taking into account the uncertainty in $b$ by
performing an average of p-values weighted by the Bayesian
posterior $p(b|y)$, that is

$p_{Ba} = \int p_P(\geq x|b)\  p(b|y)\  d\,b$\\ This can be
evaluated by Monte Carlo integration, or by a mixture of
analytical and numerical methods.  I will pursue the latter
course here.  The most common usage in HEP is to represent
$p(b|y)$ as a truncated normal distribution

$p_N (b|y) = \frac{1}{\delta b \sqrt{2 \pi}} \mathrm{exp}
\frac{-(b - \alpha y)^2}{2 (\delta b)^2}, \ b > 0\ .$\\
If $b$ is a sum of many contributions, its distribution should
asymptotically approach a normal.  An alternative I have
advocated in HEP\cite{jim}, and which is also known to the GRA
communitity\cite{alex}, is to start from a flat prior for $b$ and
derive the $p(b|y)$ in the usual Bayesian fashion, leading to a
Gamma posterior:

$p_\Gamma (b|y) = \beta^y e^{-\beta}/y!\ ,\  \beta = b/\alpha$.\\
This is most appropriate when a single contribution to $b$
dominates and its uncertainty is actually due to counting
statistics. I will refer to the Z-values which result from these
two choices as $Z_N$ for the normal posterior, and $Z_{\Gamma}$
for the Gamma function posterior.  Choosing to represent $p_P$ as
a sum, and performing the $b$ integration first gives the p-value
for the Gamma posterior\cite{alex}

$p_{\Gamma} = \sum_{j=x}^{\infty}\frac{(y+j)!}{j!y!}
\frac{\alpha^j}{(1+\alpha)^{1+y+j}}$\\
Despite appearances, $p_{\Gamma}$ is identical to $p_{Bi}$. The
Beta function representation of $p_{Bi}$ is much more suitable
for large values of $x,\ y$.  The two expressions can be made
somewhat closer by using $w = \alpha/(1+\alpha)$.

 Bayesian
practice  typically focuses on direct comparison of specific
hypotheses through the odds ratio.  However, predictive
inference\cite{gelman} is commonly used in model checking
(significance testing is just checking the background-only model).
Predictive inference in our case is directly related to
calculating $p(x|y)$, that is, averaging over the unknown
parameter $b$.

$p(j|y) = \int \ p(j|b)\  p(b|y)\  d\,b$\\
Interestingly, some Bayesian practitioners go farther, and are
willing to calculate a ``Bayesian p-value''\cite{gelman},

$p_{Bayes} = \sum_{j=x}^{\infty} p(j|x)$\\
which is precisely the $p_{Ba}$ given above (there  we summed
before integrating).

\section{COMPARISON OF RESULTS: RELATIVE PERFORMANCE}
I have taken several interesting test cases from the HEP and GRA
literature.  The input values and Z-value calculation results are
shown in Table 1.  For the HEP cases, the values reported in the
papers are $N_{on}$, $b$, and $\delta b$, while in the GRA case,
the reported values are $N_{on}$, $N_{off}$, and $\alpha$. I have
also included a few artificial cases in order to sample the
parameter space reasonably.

It is worth remarking that there are numerical issues to be faced
in evaluation of the more complex methods.  These remarks
apply--at a minimum--to a Mathematica implementation. The
Binomial is  straightforward in its Beta function
representation.  The Bayes p-value methods may involve an infinite
sum, and are touchy and slow for large $n$;  \cite{alex} suggests
approximating the summation by an integral. Fraser-Reid and the
Bayes p-value summation results may be sensitive to whether
integers are floating point values are used. An alternative
attack is to leave the $p_P$ as a $\Gamma$ function ratio and
trade an integration for the infinite sum. Doing so in the Bayes
Gaussian case is less unstable than summing, but for large $n$
requires hints on the location of the peak of the integrand.

For the purposes of the present section, I will take the
Frequentist UMPU Binomial ratio test as a reference standard,
because of its optimality properties.  I will have more to say on
this later.

None of these examples from the recent literature was published
with a seriously wrong significance level.  To me, the most
striking result in the table is that the Bayes Gamma prior method
produces results \textit{identical} to the Binomial result (MSU
graduate student HyeongKwan Kim has proven the identity).

The method most used in HEP, Bayes with a normal posterior for b,
produces Z's always larger than those from Bayes Gamma. Viewing
the calculation as averaging the Poisson p-value $p_P(b)$ over the
posterior for $b$, the shorter tails of the normal compared to
the gamma place less weight on the larger probabilities (smaller
p-values) obtained when the off-source measurement happens to
underestimate the true value of b. The difference is most
striking for large values of $\alpha$, that is, when the
background estimate is performed with less sensitivity than the
signal estimate; in this case, results differing in significance
by over .5 $\sigma$ can occur. The most common method in GRA, the
simple Log Likelihood ratio formula, produces comparable or
slightly higher estimates of significance, but seems less
vulnerable to problems at large $\alpha$.  It appears to claim
the highest significance of these methods at small $n$. The
variance stabilization method $Z_0$ presented in \cite{zhang}
does not appear to be in general use in GRA, but produces results
of similar quality to the other two mainstay methods.  All
methods agree for $N>500$, where the normal approximations are
 good, even out to 3-6 $\sigma$ tails.

The ``not recommended'' methods all produce results off by more
than .5 $\sigma$ for several low-statistics cases. $Z_9$, which
approximates $Z_{Bi}$, does best; $Z_5$ is indeed biased against
real signals compared to other measures, and its alleged
improvement $Z_5\prime$, while curing that problem, overestimates
significance as the price for its less efficient use of
information compared to $Z_9$.

As expected, ignoring the uncertainty in the background estimate
leads to overestimates of the significance.  $s/\sqrt{b}$ is much
more over-optimistic than an exact Poisson calculation,
particularly for small $n$, or $\alpha > 1$, where the background
uncertainty is most important. The best that can be said for
$s/\sqrt{b}$ is that it is mostly monotonic in the true
significance, at least as it is typically used (for comparing two
selection criteria with N varying by an order of magnitude at
most).  The 3rd order Fraser-Reid approximation is fast and
accurate up to moderate $n$, suggesting it is worth pursuing the
full nuisance parameter case.  However, the approximation fails
for one large $Z$, and is very slow for the largest $n$.
% \footnote{Mathematica
%produced no result for case \cite{Milagro} after an hour on a
%modern desktop processor.}

Of the ad-hoc corrections for signal uncertainty, none are
reliable; the ``corrected'' Poisson calculation is less biased
than the un-corrected, but still widely overestimates significance
for $\alpha > 1$, and can't be used for serious work. The
$s/\sqrt{b+\delta b}$ isn't  much better than its
``un-corrected'' version.

To summarize, most bad formulae overestimate significance (the
only exceptions are $Z_5$ for $\alpha < 1$ and Poisson with $b
\rightarrow b + \delta b$). Thus, prudence demands using a
formula with good properties. The Binomial test seems best for
simple Poisson backgrounds. For backgrounds with several
components, compare Bayes MC with $\Gamma$ or Normal posteriors.

\begin{table*}[t]
\begin{center}
\begin{tabular}{|l|c|c|c|c|c|c|c|c|c|c|c|c|} \hline
Reference&\cite{Alexe}&\cite{Top3}&\cite{Top1}
&\cite{Top2}&\cite{Zhangex}&\cite{Zhangex}&\cite{alarge}&\cite{Hegra}&\cite{Milagro}&
\cite{Whipple}&\cite{Milagro}&RMS\\ \hline
Non = x&4&6&9&17&50&67&200&523& 167589 & 498426 &2119449& \\
Noff = y&5&18.78&17.83&40.11&55&15&10&2327& 1864910 & 493434 &
23671193&\\
 $\alpha$&0.2&0.0692&0.2132&0.0947&0.5&2.0&10.0&0.167&0.0891&1.000&0.0891&\\
$b = \alpha y$&1.0&1.3&3.8&3.8&27.5&30.0&100.0&388.6& 166213 &
493434
&2109732 &\\
 s = Non -
b&3.0&4.7&5.2&13.2&22.5&37&100&134.4&1376&4992&9717&\\
$
\delta b$&0.45&0.3&0.9&0.6&3.71&7.75&31.6&8.1& 121.7 & 702.4 & 433.6& \\
$\delta b$/b&0.447&0.231&0.237&0.158&0.135&0.258&0.316&0.0207&0.000732&0.00142&0.000206&\\
Reported p&&.0030&.027&2.0E-06&&&&&&&&\\
%Reported p&&$3\times 10^{-3}$&$2.7\times 10^{-2}$&$2\times10^{-6}$&&&&&&&& \\
Reported Z&&2.7&1.9&4.6&3.0&3.0&&5.9&3.2&5.0&6.4&\\ \hline

 Recommended:&&&&&&&&&&&&\\
$Z_{Bi}\ \ $Binomial&\bf 1.66&\bf 2.63&\bf 1.82&\bf 4.46&\bf
2.93&\bf 2.89&\bf 2.20&\bf 5.93&\bf 3.23&\bf 5.01&\bf
6.40&0\\
$Z_\Gamma\ \ \ $Bayes Gamma &\bf 1.66&\bf 2.63&\bf 1.82&\bf
4.46&\bf 2.93&\bf 2.89&\bf 2.20&\bf 5.93&*&*&*&0\\
\hline

Reasonable:&&&&&&&&&&&&\\

$Z_N\ \ $Bayes Gauss (HEP)
&1.88&2.71&1.94&4.55&3.08&\it 3.44&\it 2.90&\bf 5.93&\bf 3.23&\bf 5.02&\bf 6.40&.28\\
$Z_0\ \ \sqrt{} $ +
3/8&1.93&2.66&1.98&4.22&3.00&3.07&2.39&5.86&\bf 3.23&\bf 5.01&\bf 6.40&.15\\
$Z_L\ \ $L Ratio
(GRA)&1.95&2.81&1.99&4.57&3.02&3.04&2.38&\bf 5.93&\bf 3.23&\bf 5.01&\bf 6.41&.14\\
\hline
Not Recommended:&&&&&&&&&&&&\\
$Z_9=s/\sqrt{\alpha (N_{on} + N_{off})} $&\it 2.24&\it 3.59&2.17&\it 5.67&3.11&\bf 2.89&\bf 2.18&6.16&\bf 3.23&\bf 5.01&\bf 6.41&.52\\
$Z_5 = s/\sqrt{N_{on} + \alpha ^2 N_{off}}$&1.46&\it 1.90&1.66&\it 3.17&2.82&3.28&\it 2.89&\it 5.54&\bf 3.22&\bf 5.01&\bf 6.40&.93\\
$Z_5\prime = s/\sqrt{ \alpha(1+\alpha)N_{off}}$&\it 2.74&\it 3.99&\it 2.42&\it 6.47&\it 3.50&\it 3.90&\it 3.02&6.31&\bf 3.23&\bf 5.03&\bf 6.41&.53\\

\hline
Ignore $\delta b$:&&&&&&&&&&&&\\
 $Z_P\ \ $Poisson: ignore
$\delta b$&2.08&2.84&2.14&4.87&\it 3.80&\it 5.76&\it 7.72&\it 6.44&3.37&\it 7.09&6.69&1.9\\
$Z_3\ $Fraser-Reid $\approx
Z_P$&2.07&2.84&2.14&4.87&\it3.80&\it5.76&\it(8.95)&\it6.44&3.37&6.09&6.69&2.2
\\
$Z_{sb}\ \  =s/\sqrt{b}$&\it 3.00&\it 4.12&\it 2.67&\it 6.77&\it 4.29&\it 6.76&\it 10.00&\it 6.82&3.38&\it 7.11&6.69&2.9\\
 \hline
 Unsuccessful Hacks:&&&&&&&&&&&&\\

Poisson: Nb $\rightarrow$  b +
$\delta b$&1.56&2.46&1.64&\bf 4.47&3.04&\it 4.24&\it 5.51&6.01&3.07&\it 6.09&\bf 6.39&1.1\\
s / $\sqrt{b + \delta b}$
&\it 2.49&\it 3.72&\it 2.40&\it 6.29&\it 4.03&\it 6.02&\it 8.72&\it 6.75&3.37&\it 7.10&6.69&2.4\\

\hline
\end{tabular}
\label{results} \caption{: Test Cases and Significance Results:
Inputs are at top; $\alpha$ deduced from Eq.1 for HEP examples.
The test cases are ordered in data counts; \protect\cite{Zhangex};
\protect\cite{alarge}, and \protect\cite{Whipple} have large
values of $\alpha$, troublesome for some methods. Z-values in
\textbf{bold} are nearly equal the Binomial values; Z-values in
\textit{italics} differ by more than .5 . * indicates convergence
failure. The last column gives the un-weighted RMS difference of
the Z-values from to the Binomial values.}
\end{center}
\end{table*}

% tables should appear as floats within the text
% The ruledtabular enviroment adds doubled rules to table and sets a
% reasonable default table settings.
% Use the table* environment to get a full-width table in two-column
% Add \usepackage{longtable} and the longtable (or longtable*}
% environment for nicely formatted long tables. Or use the the [H]
% placement option to break a long table (with less control than
% \begin{table}%[H] add [H] placement to break table across pages
% \begin{ruledtabular}
% Lines of table here ending with \\
% \end{tabular}
% \end{ruledtabular}
% \end{table}

% Surround table environment with turnpage environment for landscape
% table
% \begin{table}
% \begin{ruledtabular}
% \end{tabular}
% \end{ruledtabular}
% \end{table}

%\begin{table}[t]

%\begin{figure}
%\includegraphics[width=65mm]{JACpic_mc.eps}
%\caption{Layout of papers.}
%\label{l2ea4-f1}
%\end{figure}

\section{CALIBRATION OF ABSOLUTE SIGNIFICANCE: MONTE CARLO}
In the previous section, results of significance calculations
were compared to a reference calculation, the UMPU Binomial
Test.  That method produces the lowest reported significance
among the methods with a sound theoretical basis. This alone
could justify its use (on grounds of conservatism)\cite{zhang},
but would beg the question of whether the Binomial test is
actually ``correct.'' This has been studied by Monte Carlo
simulation\footnote{There may have been typographical errors in
the results for $Z_{Bi}$, identical to $Z_9$, but described as
having different deviations from the true MC result. If the Z's
were, by coincidence, identical, this might be an instance of the
measure-dependence described below. Alas, the paper was published
without the MC comparisons figures.} in \cite{zhang}.

A few observations on MC testing are useful. One might imagine
simply generating instances of Poisson variables $x, y$ with
means $\mu,\ \mu/\alpha$, and calculating $Z^{MC}$ from $p^{MC}
=$ the fraction of events ``more signal-like'' than
$(N_{on},N_{off})$. Instead, \cite{li,zhang} a separate MC is done
\textit{for each individual measure}, because there is no unique
``correct'' Z-value for a given observation.  The best that can
be done is to ask that a method produce a Z value consistent with
MC probabilities when the observation is analyzed by that method.
The problem is that there is no unique definition of ``more
signal-like''. One is essentially trying to find a unique
ordering of points on the $x,y$ plane to define those which are
similarly far from the observed point $N_{on},N_{off}$.

Each variable introduces its own metric, and contours of equal
$Z$ do not coincide for different $Z$ variables, as seen in
Figure 1.

\begin{figure}[t]
  \begin{center}
  \begin{tabular}{cc}
  \includegraphics*[height=40mm]{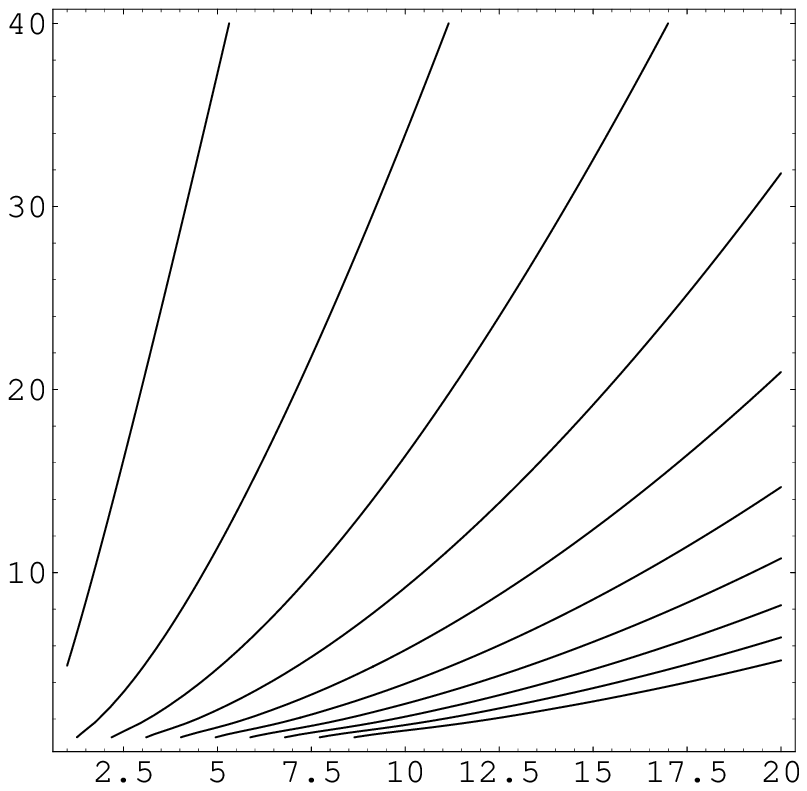} &
    \includegraphics*[height=40mm]{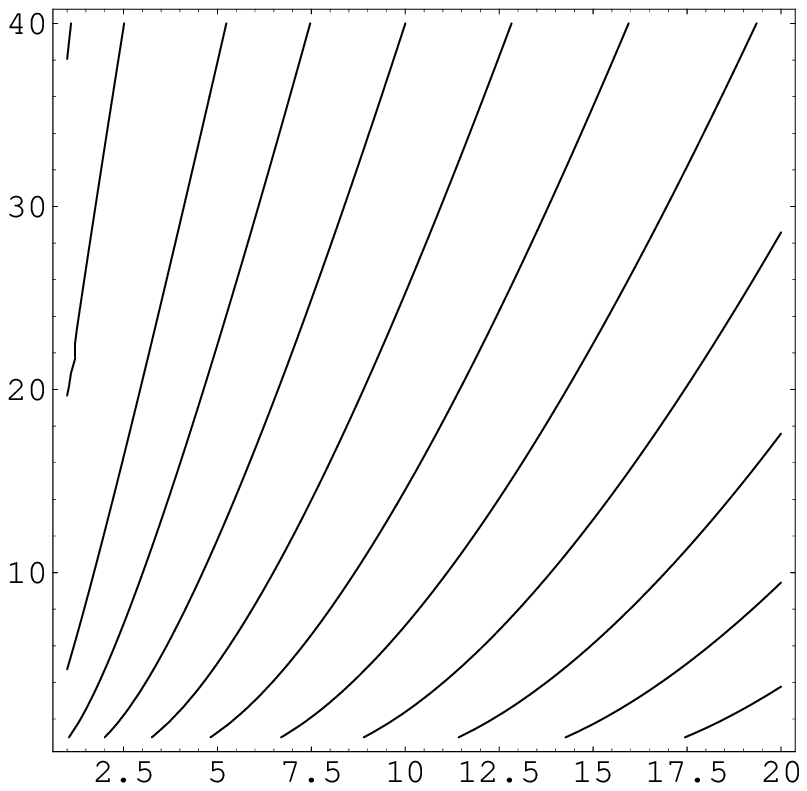}
  \end{tabular}
  \caption{Contours of equal Z, case \cite{Top2}, for $Z_{SB}$ (left) and $Z_L$ (right).}
  \label{contours}
  \end{center}
  \end{figure}

The p-value for an observation $(x_0,y_0)$ depends on these
contours:

$p^{MC}(x_0,y_0) = \int_{Z>Z_0} p(x,y)\ dx \ dy$\\
where the integration is over the region beyond the contour line
$Z_0$ passing through the observation: $Z(x,y)
> Z_0(x_0,y_0)$.

For small $n$, the contours are markedly different, so that two
\textit{different} Z-values could both be correct if each agreed
with their respective $Z^{MC}$.  Still, the situation is not
catastrophic, as values of $Z$ are not wildly different, and
presumably the $Z^{MC}$ differ somewhat less than the reported
values in Table 1. For larger $n$, the contours become straighter
and more similar, and more importantly, the probability becomes
more peaked, so that a smaller region contributes.  Thus, the
central limit forces convergence to a unique $Z$ value for large
$n$.

Although Monte Carlo studies can never explore the entire
parameter space, the general conclusion of \cite{zhang} is that
$Z_{Bi}$ is the best of the alternatives.  $Z_{Bi}$ is only
slightly conservative for $Z>3$.  There, $p_{Bi}$ is a bit larger
than $p^{MC}$ and thus $Z_{Bi} < Z^{MC}$  by  $3 \%$ or less on
the Z scale when $\mathrm{min}(N_{on},N_{off}) < 20$, and
$Z_{Bi}$ performs even better for larger $n$.   They found the
deviations of other methods from $Z^{MC}$ are typically larger.
They also cite work\cite{dag} which finds larger fractional
deviations\footnote{  It is not clear whether these limitations
(originally studied in the purely-binomial setting) are due to
discreteness; or whether the conditioning on $N_{on}+N_{off}$
causes the differences from Monte Carlo.} for $Z_{Bi}$ for
smaller Z. Since $Z>3$ is the lower edge of the region where
claims are liable to be made, and the degree of conservatism is
small, this would also justify accepting $Z_{Bi}$ as the
reference standard, and as the recommended method of evaluating
significance when there is any concern about the validity of
other methods--at least when a single counting uncertainty
dominates the knowledge of the background.

\begin{acknowledgments}
The author wish to thank LANL for hospitality and financial
support during his sabbatical,  Milagro and D0 colleagues for
information and references, and Tom Loredo for reference
\cite{zhang}. This work was supported in part by NSF contract
NSF0140106.  The
 calculations were performed with the
assistance (mostly) of \textit{Mathematica}.
\end{acknowledgments}


\end{document}